\documentclass[12pt,twoside]{article}
\usepackage{fleqn,espcrc1}

\usepackage{graphicx}

\newcommand{\Red}{}
\newcommand{\Green}{}

\def\vk{\vec k_\perp}

\newcommand{\nn}{\nonumber}
\newcommand{\wt}{\widetilde}
\newcommand{\dg}{\dagger}
\newcommand{\ck}{\check}
\renewcommand{\d}{{\rm d}}
\newcommand{\g}{\gamma}

\newcommand{\me}[3]{\langle#1\vert\,#2\,\vert#3\rangle}
\newcommand{\ovl}{\overline}

\title{Connecting generalized parton distributions and light-cone wave 
functions}

\author{Rainer Jakob\address{Fachbereich Physik, Universit\"at Wuppertal, 42097 Wuppertal,Germany}}
       
\begin{document}

\maketitle

\begin{abstract}
The relation of generalized (skewed) quark distributions to nucleon wave
functions is discussed in the context of light-cone quantization.
\end{abstract}

\section{INTRODUCTION}

Hard (semi-)inclusive processes, like deep inelastic lepton-nucleon scattering,
the Drell-Yan process, or $W$-production have revealed a lot of the hadronic
substructure in the last three decades. When the hard scale of the process 
becomes large enough, the resolution is sufficient to probe structures 
down to fractions of femtometer, i.e.~one is sensitive to the substructure of 
hadrons in terms of quarks and gluons. The interpretation of the hard 
reactions in the context of the (QCD improved) parton model relies on the 
concept of factorization: The hard process-dependent parts are calculated 
according to the rules of perturbative QCD, and the soft process-independent 
parts, which describe properties of the hadrons involved  in the reaction, are 
parameterized as soft functions, parton distribution and fragmentation 
functions. Those soft 
functions are universal, i.e.~once determined in a hard reaction they can 
be used in the same form as input to predict any other hard reaction. A 
strict formal definition as hadronic matrix elements of quark and gluon 
field operators can be given for the soft functions and their logarithmic 
scale dependence is well-understood in terms of the DGLAP evolution.

A simple probabilistic interpretation of parton distribution and fragmentation
function arises within the context of light-cone quantization. Actually, those
functions contain information on the structure of hadrons as seen by highly
relativistic probes, the distribution functions are probabilities for the
quanta of the independent dynamical fields, the so-called `good' components 
of quark and gluon fields. The operator combinations in the definitions
of leading twist soft functions are bilocal, the distance between the
arguments of the field operators is light-like. 

Another type of hadronic matrix elements of quark fields is involved in the
description of exclusive reactions. Elastic form factors, or transition form
factors are defined via matrix elements of currents expressed in terms of the
elementary (quark) fields between different hadronic initial and final 
states. The combination of field operators here is local, but the momenta
characterizing the initial and final states are different. 
   
A generalization of the above described types of hadronic matrix elements
is involved in the description of exclusive reactions like Compton scattering
and deeply virtual (hard) meson production. The operators contributing at
leading order in a twist expansion are bilocal (with a light-like distance)
and the matrix elements are off-forward (or non-diagonal) in initial and final
states. The functions parameterizing the matrix elements are called `Skewed
Parton Distributions' (SPDs).

Much theoretical interest was turned to the investigation of SPDs, since 
they provide links between inclusive and exclusive 
quantities~\cite{Dittes:1988xz,Muller:1994fv,Ji:1997nm,Ji:1998pc,Radyushkin:1996nd,Radyushkin:1996ru,Radyushkin:1997ki,Collins:1997fb,Ji:1998xh,Collins:1999be}.
Moreover, 
skewed parton distributions give access to the angular momentum of partons 
inside hadrons as was first recognized by Ji~\cite{Ji:1997ek} .

In this contribution the connection of quark SPDs to another fundamental 
quantity, the light-cone wave function of the nucleon, which describes 
how the nucleon is built up from partons in a specific configuration, is 
discussed.

\section{GENERALIZED (SKEWED) PARTON DISTRIBUTIONS}

\begin{figure} 
\begin{center} 
\includegraphics[width=9truecm]{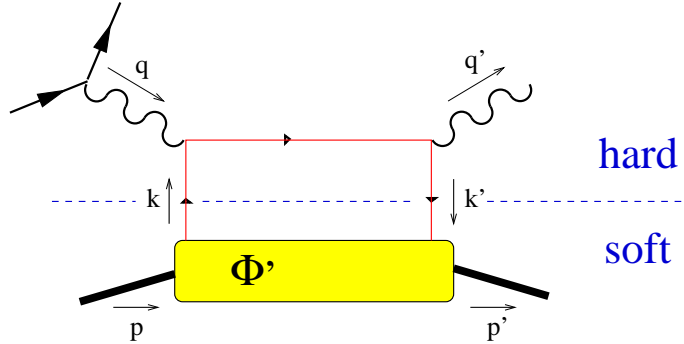}
\end{center} 
\caption{Handbag diagram contributing dominantly to Compton scattering in the
  deeply virtual region. The lower soft part consists of a hadronic matrix 
  element which is parameterized in form of skewed parton distribution
  functions.} 
\label{fig:one}
\end{figure} 
  
Skewed parton distributions are a new tool for the investigation of hadronic 
substructure. They are closely related to the ordinary (forward) parton
distributions and to form factors. To make the relationship evident it is most
easy to start from the operator definition of quark SPDs
\begin{eqnarray}
\lefteqn{p^+\!\int{dz^-\over 2\pi}\;e^{ik^+z^-}
\me{\Green{p^\prime,s^\prime}}
   {\ovl\psi(\Red{0})\,\left\{\g^+\atop\g^+\g_5\right\}\,\psi(\Red{z^-})}
   {\Green{p,s}}}\nn\\[2mm]
&&=\bar u(p^\prime,s^\prime)
\left\{\displaystyle
\g^+\;\;\fbox{\mbox{${\wt{\cal F}_\zeta(x,t)}$}}
{}+{i\sigma^{+\nu}\Delta_\nu\over 2M}\;\;
   \fbox{\mbox{${\wt{\cal K}_\zeta(x,t)}$}}
\atop\displaystyle
\g^+\,\g_5\;\;\fbox{\mbox{${\wt{\cal G}_\zeta(x,t)}$}}
{}+{\Delta^+\over 2M}\;\;
   \fbox{\mbox{${\wt{\cal L}_\zeta(x,t)}$}}
\right\}u(p,s)\;.   
\end{eqnarray} 
The notation for the momenta involved is indicated in Fig.\ref{fig:one}; the 
momentum transfer is denoted $\Delta=p'-p$. We
use light-cone components of four-vectors defined as 
$a^\pm=(a^0\pm a^3)/\sqrt{2}$ for an arbitrary vector $a^\mu$, and the 
argument of the second quark field is a shorthand for the point
$(0,z^-,\vec 0_\perp)$ on the light-cone. The SPDs depend on the fractional 
momentum of the emitted quark $x=k^+/p^+$, the `skewedness' parameter 
$\zeta=-\Delta^+/p^+$ which denotes the difference of momentum fractions 
on the two quark lines, and
the invariant momentum transfer squared $t=\Delta^2$. The above definition 
has to be
compared with the one for the (polarized) quark distribution function in a 
nucleon
\begin{equation}
p^+\!\int{dz^-\over 2\pi}\;e^{i\,k^+ z^-}
\me{{p,s}}{\ovl\psi({0})\,\left\{\g^+\atop\g^+\g_5\right\}\,\psi({z^-})}{{p,s}}
=\bar u(p,s)\,\left\{\g^+\;\;\fbox{\mbox{$q(x)$}}\atop
                     \g^+\g_5\;\;\fbox{\mbox{$\Delta q(x)$}}\right\}\,u(p,s) 
\;.
\end{equation} 
In the forward case, i.e.~for a matrix element diagonal in the nucleon
momenta, there are no analogues of the helicity-flip terms 
$\wt{\cal K}_\zeta(x,t)$ and $\wt{\cal L}_\zeta(x,t)$. In 
contrast, helicity-flip terms show up in the definition of the electromagnetic 
nucleon form factors
\begin{equation} 
\me{{p^\prime,s^\prime}}
{\ovl\psi({0})\,\left\{\g^+\atop\g^+\g_5\right\}\,\psi({0})}{{p,s}}
=\bar u(p^\prime,s^\prime)
\left\{\displaystyle
\g^+\;\;\fbox{\mbox{${F_1(t)}$}}
  {}+{i\sigma^{+\nu}\Delta_\nu\over 2M}\;\;\fbox{\mbox{${F_2(t)}$}}
\atop\displaystyle
\g^+\,\g_5\;\;\fbox{\mbox{${G_A(t)}$}}
  {}+{\Delta^+\over 2M}\;\;\fbox{\mbox{${G_{P}(t)}$}}
\right\}
u(p,s)  \;,
\end{equation} 
where initial and final nucleon momenta are different. The form 
factors $F_1(t)$, $F_2(t)$, $G_A(t)$, and $G_P(t)$ are the Dirac,
Pauli, the axial and the pseudoscalar form factors, respectively. By 
comparison reduction formulas can be read off as
\[
\lim_{t\to 0}\;
\left\{{\wt{\cal F}_\zeta(x,t)}\atop{\wt{\cal K}_\zeta(x,t)}\right\}
=\left\{q(x)\atop \Delta q(x)  \;, \right.
\]
where the formal forward limit $t\to 0$ implies $\zeta\to 0$. The lowest
moments in $x$, from which the $\zeta$ dependence drops, relate the SPDs to 
form factors
\[
\int_{-1+\zeta}^1 dx \;
\left\{{\wt{\cal F}_\zeta(x,t)}\atop{\wt{\cal G}_\zeta(x,t)}\right\}
=\left\{F_1(t)\atop G_A(t) \right.
\qquad
\mbox{and}
\qquad
\int_{-1+\zeta}^1 dx \;
\left\{{\wt{\cal K}_\zeta(x,t)}\atop{\wt{\cal L}_\zeta(x,t)}\right\}
=\left\{F_2(t)\atop G_P(t) \;.\right.
\]

\section{CONNECTION TO LIGHT-CONE WAVE FUNCTIONS}

\begin{figure}[htb] 
\begin{center} 

\includegraphics[width=6cm]{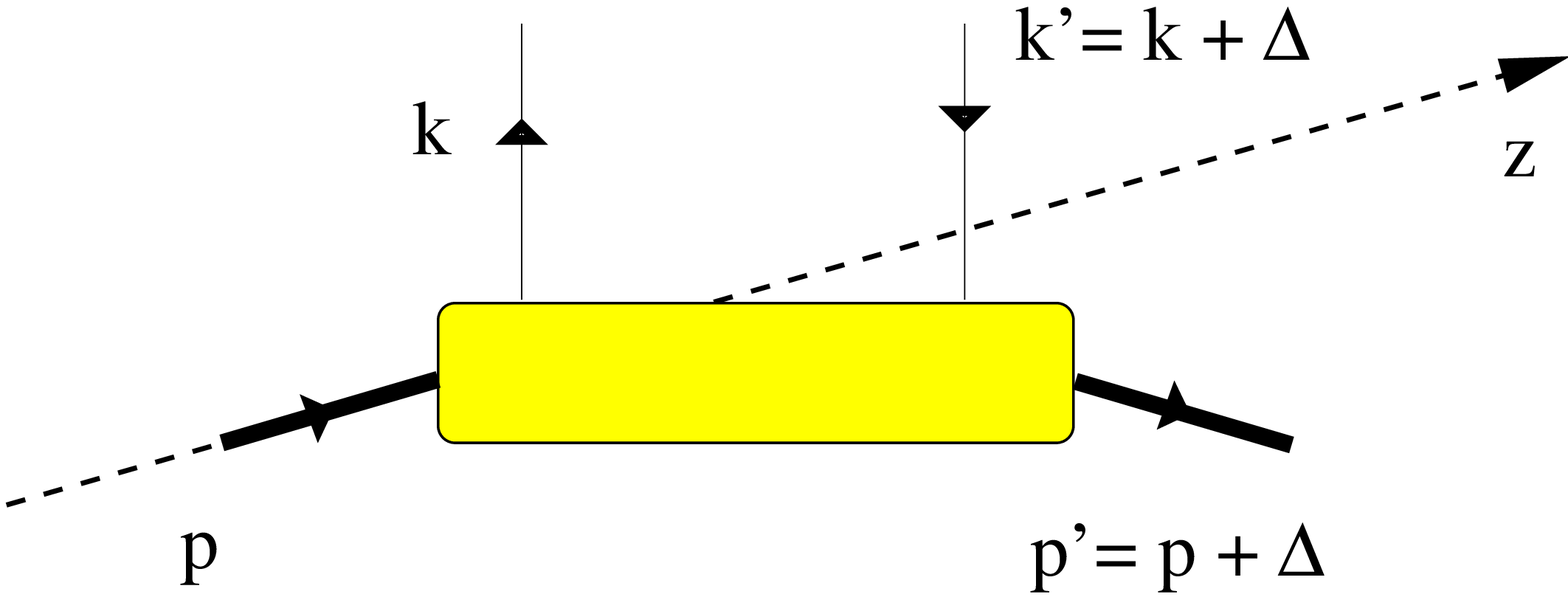}
\hspace{16mm}
\includegraphics[width=6cm]{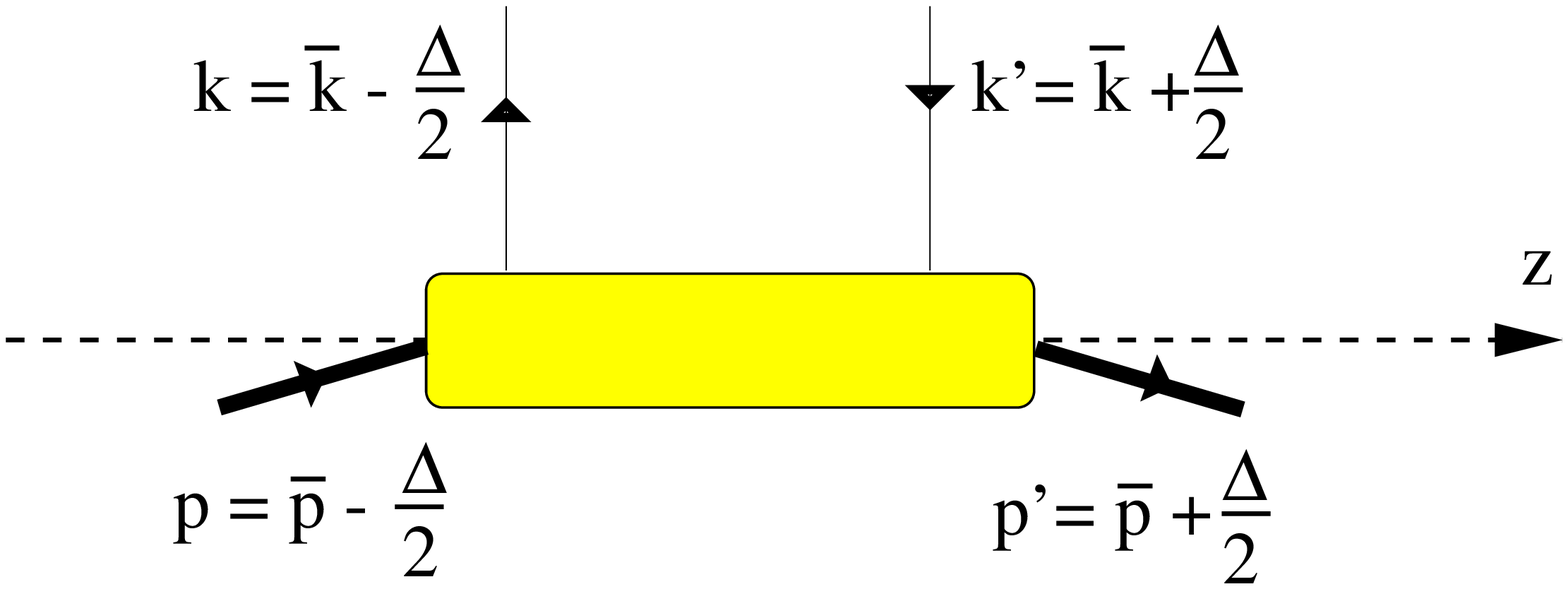}\\

\unitlength=0.5truemm
\begin{picture}(100,20) 
\put(-10,10){\line(1,0){120}}
\put(100,10){\vector(1,0){10}}
\put(112,8){$x$}
\thicklines
\put(0,9){\line(0,1){3}}\put(-16,2){$-1+\zeta$}
\put(40,9){\line(0,1){3}}\put(39,2){$0$}
\put(60,9){\line(0,1){3}}\put(59,2){$\zeta$}
\put(100,9){\line(0,1){3}}\put(98,2){$1$}
{\multiput(0,8.5)(4,0){10}{\line(1,1){3}}}
{\multiput(38.5,8.5)(4,0){5}{\line(1,1){3}}}
{\multiput(56.5,8.5)(4,0){10}{\line(1,1){3}}}
\put(2,-6){{antiquarks}}
\put(70,-6){{quarks}}
\thinlines
\put(28,70){$x$}
\put(54,70){$x-\zeta$}
\end{picture}
\hspace{26mm}
\unitlength=0.5truemm
\begin{picture}(100,20) 
\put(-10,10){\line(1,0){120}}
\put(100,10){\vector(1,0){10}}
\put(112,8){$\bar x$}
\thicklines
\put(0,9){\line(0,1){3}}\put(-4,2){$-1$}
\put(40,9){\line(0,1){3}}\put(34,2){$-\xi$}
\put(50,9){\line(0,1){2}}\put(49,2){$0$}
\put(60,9){\line(0,1){3}}\put(60,2){$\xi$}
\put(100,9){\line(0,1){3}}\put(100,2){$1$}
{\multiput(0,8.5)(4,0){10}{\line(1,1){3}}}
{\multiput(38.5,8.5)(4,0){5}{\line(1,1){3}}}
{\multiput(56.5,8.5)(4,0){10}{\line(1,1){3}}}
\put(2,-6){{antiquarks}}
\put(70,-6){{quarks}}
\thinlines
\put(20,70){$\bar x+\xi$}
\put(52,70){$\bar x+\xi$}
\end{picture}

\end{center} 
\caption{Illustration for two different common choices of reference frames. On
  the left, the longitudinal direction is defined by the initial 
  nucleon and $x$ and $x-\zeta$ are the fractional momenta of the quarks with
  respect to the initial nucleon momentum. On the right, an average of the 
  initial and final nucleon momenta defines the longitudinal
  direction, and $\bar x$ is an average of the fractional momenta the two quark
  lines carry with respect to that average nucleon momentum. Kinematical
  regions for the interpretation in terms of quarks and anti-quarks are
  indicated. The central region in both cases corresponds to the emission
  (absorption) of a quark-antiquark pair.}
\label{fig:two}
\end{figure}

Like the ordinary (forward) parton distributions also SPDs acquire a simple
interpretation as probability densities in terms of quanta of `good'
components of the fields. The `good' components of the quark fields are
projected out as $\psi_+(z)=P_+\psi(z)$ with $P_+=(\g^-\g^+)/2$ and have a 
momentum decomposition 
\begin{eqnarray} 
\psi_+^i(z^-,{\bf z}_\perp)&=&
\int\frac{\d x_i\,\d^2{\bf k}_{\perp i}}{x_i\,16\pi^3}\,\Theta(p_i^+)\;
\sum_{\mu_i}\;\nn\\
&&\times\bigg\{
     b_i(\omega_i)\;u_+(\omega_i)
  \;e^{-i\,k_i^+z^-+i\,{\bf k}_{\perp i}\cdot{\bf z}_\perp)}
{} + d_i^\dagger(\omega_i)\;v_+(\omega_i)
  \;e^{i\,k_i^+z^--i\,{\bf k}_{\perp i}\cdot{\bf z}_\perp)}
\bigg\} 
\end{eqnarray} 
where we use a collective notation for the dependence on
the plus and transverse momentum components, and on the helicity in the form
$f(x_i,{\bf k}_{\perp i}, \mu_i)=f(\omega_i)$.
The operators $b_i$ and $d_i^\dagger$ are the annihilator of the plus
component of the quark fields and the creator of the plus component of the 
antifields, respectively. They fulfill the equal light-cone 
time anticommutation relations~\cite{Brodsky:1989pv}
\begin{equation} 
\left\{b_i(\omega_i),b_j^\dagger(\omega_j')\right\}=
\left\{d_i(\omega_i),d_j^\dagger(\omega_j')\right\}=
16\pi^3\;x_i\;\delta(x_i-x_j')\;
\delta^2\left({\bf k}_{\perp i}-{\bf k}'_{\perp j}\right)\;
\delta_{\mu\mu'}\;\delta_{ij} 
\;.
\end{equation} 
The key point for a probabilistic interpretation is the observation that the
quark field operator in the definition of the hadronic matrix elements is a
density in terms of the `good' 
components, i.e.
\begin{equation}
\psi(0)\g^+\psi(z^-)=\sqrt{2}\;{\psi_+}^\dg(0)\psi_+(z^-) \;.
\end{equation} 
The quark SPD 
describes the emission of an (anti-)quark
from the nucleon with a certain momentum fraction $x$ and its subsequent
reabsorption with a different momentum fraction $x-\zeta$. In addition there 
is a kinematical region where the nucleon emits (or absorbs) a
quark-antiquark pair. 
In fixing the notations for momenta and their fractions one has to define the
longitudinal direction (i,.e.~to chose a frame of reference). Two different
popular choices are indicated in Fig.\ref{fig:two} characterized by the
momentum fractions ($x$, $\zeta$) and ($\bar x$, $\xi$), respectively. In this
contribution we adhere throughout to the first choice and follow the notations
in~\cite{Radyushkin:1997ki}. 

A connection between light-cone wave functions and form factors can be
established by assuming that the nucleon states may be replaced by a 
superposition of partonic Fock states containing 
quanta of the `good' light-cone components of (anti-)quark and gluon fields
\begin{equation}
|nucleon;p,s\rangle = \sum_{N,\beta} \;
\int [\d x]_N [\d^2 {\bf k}_\perp]_N\;
\Psi_{N,\beta}(x,{\bf k}_\perp) \;
|N,\beta\rangle \;,
\label{Fockstate}
\end{equation} 
with
\begin{equation}
[\d x]_N=\prod_{i=1}^N \d x_i \; \delta\left(1-\sum_m x_m\right)
\qquad
[\d^2 {\bf k}_\perp]_N=\prod_{i=1}^N \d^2 {\bf k}_{\perp i} \;
\delta^{(2)} \left(\sum_m{\bf k}_{\perp m}\right)
\end{equation} 
and where $\Psi_{N\beta}(x,{\bf k}_\perp)$ is the light-cone momentum wave
function of the $N$ parton Fock state. The index $\beta$ is a collective
quantum number and labels the different ways of coupling the partons into the
nucleon. Using the momentum decomposition of 
the quark field operators and the commutation relations for the creation and
annihilation operators one derives the well-known Drell-Yan
formula for the contribution of the $N$ parton Fock state to the Dirac form 
factor~\cite{Drell:1970km}
\begin{equation}
F_1^{a(N)} (t) =
      e_a\,\sum_j\sum_\beta\int[d x]_N[d^2 k_\perp]_N \;
      \Psi^*_{N,\beta}(x,\vk{}')\,\Psi_{N,\beta}(x,\vk{}) \;,
\label{eq:DY}
\end{equation} 
where the index $j$ denotes the active parton and runs over all partons of 
type $a$ with charge $e_a$. The full form factor is obtained 
by summation over all Fock
states. The shifted transverse momenta in the argument of the final state wave
function are to be taken as $\vk{}'_i = \vk{}_i-x_i\,{\vec\Delta}_\perp$ for
the spectator partons, and $\vk{}'_j = \vk{}_j+(1-x_j)\,{\vec\Delta}_\perp$
for the active quark. Note that the arguments of the wave functions are
light-cone momentum fractions and transverse momenta of the partons with 
respect to their parent nucleon momenta, which are different for initial and
final nucleon. The most convenient way to identify those arguments consists in
performing transverse boosts to reference frames where the parent hadrons have
no transverse momentum components. The shift in the transverse momenta of the
spectator partons is the result of the appropriate transverse boosts, whereas
the shift for the active quark combines the effects of absorbing the virtual
photon and the one of the transverse boost.

The contribution of the $N$ parton Fock state to the ordinary quark parton 
distributions are straightforwardly obtained as 
\begin{equation}
q^{(N)}(x) = \sum_j\sum_\beta\int[d x]_N[d^2 k_\perp]_N\;
   {\delta(x - x_j)} \;
  \Psi^*_{N,\beta}(x,\vk{})\,\Psi_{N,\beta}(x,\vk{})
\label{PDFprob}
\end{equation} 
based on the probabilistic interpretation. Given the relations (\ref{eq:DY})
and (\ref{PDFprob}) and the close relationship between SPDs, forward
distributions and form factors it is suggestive that there must be a similar
way to obtain the SPDs from light-cone wave functions. Indeed, following
along the same lines as in the derivation of the above formulas results in a
generalization of the Drell-Yan formula valid in the partonic regimes, 
i.e.~for $-1+\zeta\leq x\leq 0$ and $\zeta\leq x\leq 1$. 

\begin{equation} 
\widetilde{{\cal F}}_\zeta^{\,a\,(N)}({x;t}) = 
\sum_j\sum_\beta\int[d x]_N[d^2 k_\perp]_N\; 
{\delta(x - x_j)}\; \left(1-\zeta\right)^{{1-N} \over 2}\;
\Psi^*_{N,\beta}({\ck{x}',\ck{\vec{k}}_\perp^{\;\prime}})\,
\Psi_{N,\beta}(x,\vec k_\perp)
\label{eq:SPDoverlap}
\end{equation} 
the index $j$ running over all quarks of flavor $a$. The shifted arguments in
the final state wave function are to be taken as
\begin{eqnarray} 
\ck{x}'_i= \frac{x_i}{1-\zeta} &\quad& 
\ck{\vec{k}}_{\perp i}^{\;\prime} \;=\; 
\vk{}_i-\frac{x_i}{1-\zeta}\,{\vec\Delta}_\perp
\end{eqnarray} 
for the spectator partons and
\begin{eqnarray} 
\ck{x}'_j= \frac{x_j-\zeta}{1-\zeta} &\quad& 
\ck{\vec{k}}_{\perp j}^{\;\prime} \;=\; 
\vk{}_j+\frac{1-x_j}{1-\zeta}\,{\vec\Delta}_\perp
\end{eqnarray}
for the active quark. Equation~(\ref{eq:SPDoverlap}) was 
was obtained in \cite{Diehl:1999kh} identifying the dominant 
contributions in a diagrammatic approach. In \cite{Diehl:1999kh} and 
\cite{Diehl:1999tr} the connection between SPDs and light-cone wave functions
was exploited phenomenologically by explicitely modeling the wave function.
Different inclusive and exclusive quantities like the electromagnetic form
factors of proton and neutron, unpolarized and polarized forward parton
distributions, and cross sections of wide angle real and virtual Compton
scattering were studied in a consistent way. A detailed presentation of 
the derivation of an overlap formula for SPDs in the context of 
light-cone quantization, as briefly 
indicated in this contribution,  will be given elsewhere~\cite{Diehl:2000xz}.


\begin{thebibliography}{99}
\newcommand{\wwwspires}{http://www.slac.stanford.edu/spires/find/hep/www}

\bibitem{Dittes:1988xz}
F.~M.~Dittes, D.~Muller, D.~Robaschik, B.~Geyer and J.~Horejsi,
Phys.\ Lett.\  {\bf B209}, 325 (1988).

\bibitem{Muller:1994fv}
D.~Muller, D.~Robaschik, B.~Geyer, F.~M.~Dittes and J.~Horejsi,
Fortsch.\ Phys.\  {\bf 42}, 101 (1994)
[hep-ph/9812448].

\bibitem{Ji:1997nm}
X.~Ji,
Phys.\ Rev.\  {\bf D55}, 7114 (1997)
[hep-ph/9609381].

\bibitem{Ji:1998pc}
X.~Ji,
J.\ Phys.\ G {\bf G24}, 1181 (1998)
[hep-ph/9807358].

\bibitem{Radyushkin:1996nd}
A.~V.~Radyushkin,
Phys.\ Lett.\  {\bf B380}, 417 (1996)
[hep-ph/9604317].

\bibitem{Radyushkin:1996ru}
A.~V.~Radyushkin,
Phys.\ Lett.\  {\bf B385}, 333 (1996)
[hep-ph/9605431].

\bibitem{Radyushkin:1997ki}
A.~V.~Radyushkin,
Phys.\ Rev.\  {\bf D56}, 5524 (1997)
[hep-ph/9704207].

\bibitem{Collins:1997fb}
J.~C.~Collins, L.~Frankfurt and M.~Strikman,
Phys.\ Rev.\  {\bf D56}, 2982 (1997)
[hep-ph/9611433].

\bibitem{Ji:1998xh}
X.~Ji and J.~Osborne,
Phys.\ Rev.\  {\bf D58}, 094018 (1998)
[hep-ph/9801260].

\bibitem{Collins:1999be}
J.~C.~Collins and A.~Freund,
Phys.\ Rev.\  {\bf D59}, 074009 (1999)
[hep-ph/9801262].

\bibitem{Ji:1997ek}
X.~Ji,
Phys.\ Rev.\ Lett.\  {\bf 78}, 610 (1997)
[hep-ph/9603249].

\bibitem{Brodsky:1989pv}
S.~J.~Brodsky and G.~P.~Lepage,
SLAC-PUB-4947
{\it  IN *MUELLER, A.H. (ED.): PERTURBATIVE QUANTUM CHROMODYNAMICS* 93-240 AND
  SLAC STANFORD - SLAC-PUB-4947 (89,REC.JUL.) 149p}.

\bibitem{Drell:1970km}
S.~D.~Drell and T.~Yan,
Phys.\ Rev.\ Lett.\  {\bf 24}, 181 (1970).

\bibitem{Diehl:1999kh}
M.~Diehl, T.~Feldmann, R.~Jakob and P.~Kroll,
Eur.\ Phys.\ J.\  {\bf C8}, 409 (1999)
[hep-ph/9811253].

\bibitem{Diehl:1999tr}
M.~Diehl, T.~Feldmann, R.~Jakob and P.~Kroll,
Phys.\ Lett.\  {\bf B460}, 204 (1999)
[hep-ph/9903268].

\bibitem{Diehl:2000xz}
M.~Diehl, T.~Feldmann, R.~Jakob and P.~Kroll,
hep-ph/0009255.

\end{thebibliography}
\end{document}